\documentclass[preprint,showkeys,amsmath,amssymb,nofootinbib]{revtex4}

\usepackage{bm}

\newcommand{\al}{\alpha}
\newcommand{\be}{\beta}

\newcommand{\de}{\delta}
\newcommand{\eps}{\varepsilon}

\newcommand{\la}{\lambda}

\newcommand{\De}{\Delta}
\newcommand{\Ga}{\Gamma}

\newcommand{\Om}{\Omega}

\newcommand{\N}{\mathcal{N}}

\newcommand{\Db}{\bar{D}}

\newcommand{\ub}{\bar{u}}

\newcommand{\thb}{\bar{\theta}}
\newcommand{\ib}{\bar{i}}
\newcommand{\jb}{\bar{j}}

\newcommand{\Ct}{\tilde{C}}

\newcommand{\ft}{\tilde{f}}
\newcommand{\Pt}{\tilde{P}}

\newcommand{\ut}{\tilde{u}}

\newcommand{\pa}{\partial}
\newcommand{\pab}{\bar{\partial}}

\newcommand{\Be}{\begin{equation}}
\newcommand{\ee}{\end{equation}}

\newcommand{\half}{\frac{1}{2}}

\begin{document}


\title{A simple BRST system with quadratically constrained ghosts}
\author{Michael Chesterman}
 \email{chesterm@physics.unc.edu}
\affiliation{Department of physics, University of North Carolina,
Chapel Hill, NC 27599, USA }


\begin{abstract}
I present a toy model for the Berkovits pure spinor superparticle.
It is a $D=1$, $\N=2$ superparticle with no physical degrees of
freedom. We study the cohomology in various ways, in particular
finding an explicit expression for the `b'-field. Finally, we
construct the topological string B-model from a straightforward
generalization of the system.
\end{abstract}



\maketitle

\section{Introduction} \label{sec:introduction}

The pure spinor string \cite{Berkovits:2000fe}, introduced by
Berkovits in the year 2000, is the first string model to be
quantized covariantly with respect to the full super-Poincar\'e
symmetry group of flat (super)space-time. See
\cite{Berkovits:2002zk} for an early review article. It is also
arguably the most promising model for calculating higher loop
amplitudes
\cite{Berkovits:2004px,Berkovits:2005df,Berkovits:2006bk,Stahn:2007uw},
and for covariant quantization in Ramond-Ramond backgrounds
\cite{Berkovits:2004xu}.

Characterized by a free field world-sheet action and a BRST
charge, it has one unconventional feature: its bosonic ghost
fields obey certain quadratic constraints. This has interesting
implications for the BRST cohomology, which defines physical
states and operators. It also leads to some difficulties. Firstly,
a clean expression for the `b'-field, which is important in
amplitude computations, is not known in the `minimal' formalism
\cite{Berkovits:2004px}. Secondly, a covariant BRST treatment of
the quadratic constraints requires infinitely many
ghosts-for-ghosts \cite{Chesterman:2002ey,Berkovits:2005hy}. These
issues apply equally to the more straightforward superparticle,
which we will consider from now on.

To begin with, let us compare the pure spinor superparticle to a
similar but ordinary Hamiltonian BRST system, without the
constrained ghosts. See \cite{Henneaux:1992ig} for a review of
such systems.

Consider a generic abelian gauge theory, whose operators act on
some superspace with co-ordinates $X,\theta$. Suppose that it has
independent fermionic gauge generators, or first class constraints
$\{g_\al\},$ $\al =1,...,N$. We introduce bosonic ghost variables
$l^\al$, such that the local gauge symmetry is replaced with a
global symmetry defined by fermionic, nilpotent BRST charge
\begin{eqnarray}
\Om = l^\al g_\al,  && \Om^2 = 0.
\end{eqnarray}
A physical state $\psi$ is in the cohomology of $\Om$ at
ghost-number 0
\begin{equation}
g_\al \psi(X,\theta) =0.
\end{equation}
We see that $\psi$ is simply a physical wavefunction in the Dirac
quantization of the gauge theory.
At higher ghost numbers, the cohomology vanishes (at least  locally), as can be shown by making
a (local) canonical change of basis such that $g_{\al}$ are the first $N$ momenta.

Now let us consider the $\N=1$ Berkovits
superparticle\cite{Berkovits:2001rb}. Its gauge theory origin is
currently unclear, but the BRST charge is defined to be
\begin{eqnarray}
\Om = \la^\al D_\al, && \Om^2 = -i \la^\al \Ga^a_{\al\be}\la^\be
\frac{\pa}{\pa X^a},
\end{eqnarray}
where $D_\al$ is a fermionic covariant derivative which acts on
a $D=10$, $\N=1$ superspace with co-ordinates $X^a,\theta^\al$,
where $a=0,...,9$, and $\al=1,...,16$. The variables $\la^\al$ are
bosonic ghosts, which are made to obey the so-called pure spinor
constraints
\begin{eqnarray}
\la^\al\Ga_{\al\be}^a\la^\be=0,
\end{eqnarray}
in order that $\Om$ be nilpotent. See
\cite{Berkovits:2001rb,Chesterman:2004xt} for more details. We
note that the Berkovits BRST charge resembles that of the earlier
abelian gauge theory. However, constraints $D_\al$ are second
class and not abelian, which is compensated for by the quadratic
pure spinor constraints. Also, the state cohomology turns out to be non-zero at
ghost numbers zero to three.




In this article, we build and analyze a toy model, which has
analogous features to the pure spinor superparticle, but which is
simpler to handle. The aim is to help better understand the harder
properties of the string, by constructing a clean system, where
everything relating to the ghost constraints is easy to solve. In
particular, there are only finitely many ghosts and we are able to
find an explicit expression for the `b'-field, without use of picture changing operators \cite{Berkovits:2004px}, or adding non-minimal fields in the style of \cite{Berkovits:2005bt}.


The paper is structured as follows: First we construct the simple
system and calculate its spectrum. Next we calculate the operator
cohomology, which includes finding the `b'-field, and we make a
BRST implementation of the ghost constraint. We then calculate a
partition function, and show equivalence to a simple bosonic
model. Finally, as an exercise, we use our toy system to construct the topological
string B-model.

After submitting this article to arXiv.org, P.A Grassi pointed out to me some overlap with earlier work \cite{Grassi:2005jz,Adam:2006bt,Grassi:2006wh,Grassi:2007va,Grassi:2008yb}, in which he and collaborators had made use of a similar but not identical toy model. However, the main result of this paper, which is the expression for the `b'-field is not part of the overlap.

\section{The Toy Model}

The toy model is a supersymmetric quantum particle with target
space given by flat $D=1$, $\N=2$ superspace co-ordinates $X$ and
$\theta^a$, supplemented by bosonic ghost-number one variables
$u^a$, where $a=1,2$. Variables $\theta^a$ and $u^a$, are vectors
of an SO(2) R symmetry. Indices are raised and lowered with a kronecker delta.

The BRST charge is defined as
\begin{eqnarray}
\Om=u^a D_a,
\end{eqnarray}
where
\begin{eqnarray}
D_a= \frac{\pa}{\pa \theta^a} - i \theta_a \frac{\pa}{\pa X}
\end{eqnarray}
are the fermionic covariant derivatives, and the ghosts also obey
a quadratic constraint
\begin{eqnarray}
u^a u_a=0,
\end{eqnarray}
which describes a one-dimensional complex surface.
By construction, the quadratic constraint is the minimum required
such that $\Om$ be nilpotent
\begin{eqnarray}
\Om^2 = -i u^a u_a \frac{\pa}{\pa X}.
\end{eqnarray}
The above combination of quadratic constraint and superspace
describes the simplest non-trivial such system.

There is a gauge symmetry associated with quadratic constraint:
\begin{equation}\label{eq:gauge_symmetry}
\de_\eps \frac{\pa}{\pa u^a} = [\eps(\tau) \frac{\pa}{\pa u^a},
u^b u_b] = 2\eps(\tau) u_a,
\end{equation}
for small local parameter $\eps$ on the world-line $\tau$. It is
this which kills one degree of freedom of the $u^a$ momenta.

We define a BRST and gauge invariant Hamiltonian
\begin{equation}
H= \frac{1}{2} \left(\frac{1}{i} \frac{\pa}{\pa X} \right)^2,
\end{equation}
which is shown later to be BRST-exact, as is typical of particle
theories which come from string theories.

The global Poincar\'e symmetry generators
\begin{eqnarray}
\frac{\pa}{\pa X}, && Q_a=\frac{\pa}{\pa \theta^a} + i \theta_a
\frac{\pa}{\pa X}
\end{eqnarray}
(anti-)commute with the BRST charge and Hamiltonian.

\section{State Cohomology}
\subsection{Zero momentum cohomology}
It is instructive to calculate the zero-momentum cohomology. At
zero momentum the BRST charge becomes
\begin{eqnarray}
\Om_0 = u^a \frac{\pa}{\pa \theta^a}, && \Om_0^2 =0
\end{eqnarray}
which is nilpotent even for unconstrained $u^a$. Based on a
technique by Berkovits \cite{Berkovits:2002zk}, we can relate the
cohomology of $\Om_0$ with constrained ghosts $H(\Om_0| u^a
u_a=0)$ to that with unconstrained ghosts $H(\Om_0)$.

A BRST-closed wavefunction $F_0(u,\theta)$ obeys
\begin{eqnarray}
\Om_0 F_0 = u^a u_a F_1, && \Om_0 F_1 = 0,
\end{eqnarray}
for some function $F_1(u,\theta)$. The $F$'s also obey
transformations
\begin{eqnarray}
 \de F_p = \Om_0 G_p, &&
\de F_0 = u^a u_a G_1,
\end{eqnarray}
for $p=0,1$. If $F_1$ is zero, then $F_0$ belongs to the
unconstrained cohomology $H(\Om_0)$, which is known to be simply a
constant.  Otherwise, $F_1$ belongs to $H(\Om_0)$. So the $F_p$'s
give the physical content of the spectrum. Note that we are just
interested in the cohomology holomorphic in $u^a$. The
corresponding wavefunction is
\begin{equation}
\psi = C + u^a \theta_a \Ct
\end{equation}
where  $C$ and $\Ct$ are the constants corresponding to $F_0$ and
$F_1$.

In general, the above is a series of recursion relations. For the
Berkovits superparticle, the first two steps are $\Om_0 F_0 =
\la^\al \Ga^a_{\al\be} \la^{\be} F_{1a}$ and $\Om_0 F_{1a}=
\la^{\al}\Ga_{a \al\be}F_2^\be$, where $F_0$, $F_{1a}$,
$F_2^{\al}$ correspond to the ghost, gluon, and gluino of super
Yang-Mills.

So we see that in effect the quadratic constraints determine the
spin content of the physical spectrum.
\subsection{The Full Cohomology}\label{sec:state_cohomology}

An attractive property of this toy system is that the full
cohomology can be directly solved for all ghost numbers at once.
It is convenient to change to U(1) covariant co-ordinates $u=u^1
+iu^2$, $\ut=u^1-iu^2$.  Recall that $u^a$'s are complex so that
$\ut$ is not the complex conjugate of $u$. The BRST charge and
constraint become
\begin{eqnarray}
\Om = u D + \ut \Db, && u\ut=0
\end{eqnarray}
where
\begin{eqnarray}
D= \frac{\pa}{\pa\theta} -\frac{i}{2} \thb \pa_X, && \Db =
\frac{\pa}{\pa\thb} -\frac{i}{2} \theta \pa_X.
\end{eqnarray}


On the quadratic constraint surface, a general wavefunction has
the form
\begin{equation}\label{eq:wavefnc_on_uut_is_zero}
\psi = f_0(X,\theta) + \sum_{p=1}^{\infty} \left(u^p f_p(X,\theta)
+ \ut^p \ft_p(X,\theta)\right).
\end{equation}

The cohomology $H^p(\Om)$ splits into three distinct cases. When
$p \geq 2$, then
\begin{eqnarray}
f_p \in H(D)=0, && \ft_{p} \in H(\Db)=0,
\end{eqnarray}
where note that $D$ and $\Db$ are nilpotent operators. For $p=1$,
the difference to the above case, is that the variations of $f_1$
and $\ft_1$ are related
\begin{eqnarray}
\de f_1 = Dg_0 , && \de \ft_1 = \Db g_0,
\end{eqnarray}
for some function $g_0(x,\theta)$. The wavefunction takes the form
\begin{eqnarray}
\psi_1 = \half(u\thb + \ub \theta ) \Ct(X), && \de \Ct(X)=\pa_X
G(X),
\end{eqnarray}
for some function $G(X)$, where $\Ct$ has no equation of motion.
When $p=0$, $f_0$ is a constant. In other words, a physical
wavefunction takes the form
\begin{eqnarray}\label{eq:wavefnc}
\psi = C(X) + u^a \theta_a \Ct(X),
\end{eqnarray}
where
\begin{eqnarray}
\pa_X C=0, && \de \Ct(X) =\pa_X G(X).
\end{eqnarray}
$C$ and $\Ct$ resemble the ghost and anti-ghost of a
Batalin-Vilkovisky theory.



\section{Operator Cohomology}



It's straightforward to construct BRST-closed ghost number 0
operators. However, it is harder to find those which are BRST
exact. Specifically, given a BRST-exact, ghost number 0 operator
$F= [G,\Om]$, it is a non-trivial problem to find $G$, because of
the gauge symmetry associated with the quadratic constraint. The
natural ghost number $-1$ operator $\pa/\pa u^a$ is not gauge
invariant but transforms as in eqn \eqref{eq:gauge_symmetry}. So
it is not obvious how to construct negative ghost number
operators. This is the essence of the `b'-field problem of the
Berkovits superparticle. The expression for b, which is defined by
$\{b , \Om\} = P^a P_a$, where b is a regular and nilpotent operator, is not known.

Let us find a basis for the BRST-exact operators. One can think of these operators as effective constraints of the model, in the sense that they map physical states to BRST-exact states.
They must be BRST-closed, which leads to the candidates $-i\pa/\pa
X$ and $Q_a$. Using equation \eqref{eq:wavefnc}, we find that they
annihilate physical states in the following sense
\begin{eqnarray}
-i\frac{\pa}{\pa X} \psi_{\text{phys}} \sim 0, && Q_a
\psi_{\text{phys}} \sim 0,
\end{eqnarray}
where $\psi_1 \sim \psi_2$ implies that $\psi_1 = \psi_2 + \Om\phi$ for some state $\phi$.
Thus, both candidates are BRST-exact. We define the corresponding
$b$ and $h_a$ ghosts by
\begin{eqnarray}
\{b,\Om\} = -i \frac{\pa}{\pa X}, && [h_a , \Om ] =Q_a.
\end{eqnarray}
We can get a handle on $b$ and $h$ from seeing how they act on
physical states. For $b$, acting on wavefunction eqn
\eqref{eq:wavefnc_on_uut_is_zero}

\begin{eqnarray}
\Om b u^p f_p = \Om u^{p-1} \Db f_p, && \Om b \ut^p \ft_p = \Om
\ut^{p-1} D \ft_p
\end{eqnarray}
for $p \geq 1$, and $b f_0 = 0$. So we see that $b$ acts
differently on the $u=0$, $\ut=0$ and $u=\ut=0$ sectors. Thus we
define projection operator
\begin{equation}
P = \int{d \ut \de(\ut )} -\int{du}\int{d\ut}{\de(u)\de(\ut)},
\end{equation}
such that
\begin{equation}
P\psi =\sum_{p=1}^{\infty} u^p f_p(X,\theta),
\end{equation}
and an analogous operator $\Pt$.
Based on the above observations, we make an educated guess
\begin{equation}
b= u^{-1} P  \Db + \ut^{-1} \Pt D.
\end{equation}
Indeed $[b, u \ut] \approx 0$, and
\begin{equation}
\{b , \Om\} = \{ D,\Db\}\left (\int{du \de(u) + \int{d\ut \de(\ut)
- \int{du d\ut \de(u)\de(\ut)}}} \right) \approx -i\frac{\pa}{\pa
X}.
\end{equation}
Note that despite the presence of $u^{-1}$and $\ut^{-1}$ terms, $b$ is not a singular operator.

The b-field is perhaps more suggestively written as
\begin{equation}
b=\int_{C}{du^a D_a \de(u^b u_b) - \int{d^2 u \frac{u^a}{u^b u_b}D_a\de(u^1)\de(u^2) }},
\end{equation}
where $C$ is a sum of paths orthogonal to $u=0$, and $\ut=0$ respectively, and passing through the point $(u,\ut)$. One can think of
\begin{equation}
\hat{v}_a = \int_{C}du_a \de(u^b u_b) -  \int{d^2 u \frac{u_a}{u^b u_b}\de(u^1)\de(u^2) }
\end{equation}
as a gauge-invariant, ghost number -1 operator replacement for
ghost momenta $-i\pa/\pa u^a$.

Our expression is similar to that of Oda and Tonin's pure spinor
b-field \cite{Oda:2004bg,Grassi:2008yb}. To transfer their approach to our toy
model we simply replace $\hat{v}_a$ with $Y_a=(k_a/ k.u)$, where
$k_a$ is some arbitrary constant. However, a problem is that
$Y_a$ is not strictly allowed as an operator since it is singular.
One also has to find an interpretation for $k^a$, since $b$ should
be unique.


Using that $Q = D +i\thb \pa_X$, a similar analysis for h yields
\begin{equation}
h = -\thb b + \frac{\pa}{\pa u}P,
\end{equation}
\begin{equation}
[h , \Om] = Q \left (\int{du \de(u) + \int{d\ut \de(\ut) - \int{du
d\ut \de(u)\de(\ut)}}} \right) \approx Q.
\end{equation}
It will be useful to find the corresponding $b$ field expression
for the pure spinor case. Note that the $b$ and $h$ operators
aren't hermitian. We cure this in the following section by giving
a BRST treatment of the quadratic constraint.



\section{A BRST implementation of the ghost constraint}\label{sec:BRST_ghost_constraint}
When performing a path integral over the ghosts $u$ and $\ut$,
rather than doing a patch-wise integration over the constraint
surface, it makes sense to use BRST methods. It also turns out to
be useful in defining a hermition `b'-field later on. Following my
earlier work \cite{Chesterman:2002ey} for the Berkovits
superparticle, we implement the constraint with new BRST charge
\begin{eqnarray}
\De = \rho u \ut ,&& \De^2=0,
\end{eqnarray}
where $\rho$ is a fermionic ghost. $\Om$ is nilpotent and maps
between cohomology classes of $\De$ in the sense that
\begin{eqnarray}
\Om^2 = \{-i\pa_X\frac{\pa}{\pa \rho}  , \De \} \simeq 0 ,&& \{
\Om ,\De\}=0,
\end{eqnarray}
where $A \simeq B$ means that $A$ and $B$ are in the same
$\De$-equivalence class.
A physical operator $F$ is BRST-closed if
\begin{eqnarray}
[F,\De] = 0, && [F,\Om] \simeq 0,
\end{eqnarray}
and is defined up to a variation
\begin{eqnarray}
\de F \simeq [G,\Om],&& [G,\De ]=0
\end{eqnarray}
Similarly, a physical state $\psi$ obeys
\begin{eqnarray}
\De \psi = 0, && \Om\psi \simeq 0,
\end{eqnarray}
and is defined up to variation
\begin{eqnarray}
\de \psi \simeq \Om \phi, && \De\phi=0.
\end{eqnarray}
There is one ghost number operator for each BRST charge
\begin{eqnarray}
G_{\De}= \rho \frac{\pa}{\pa \rho}, && G_{\Om} = u^a
\frac{\pa}{\pa u^a} -2 \rho \frac{\pa}{\pa \rho}
\end{eqnarray}
where $G_{\Om}$ must be $\De$-closed, hence the $\rho$ dependence.
The physical wavefunction in eqn \eqref{eq:wavefnc} now appears at
$G_\De$-ghost number one. A dual wavefunction appears at ghost
number zero.

With the full BRST approach we can now write a hermitian $b$
field, given by
\begin{eqnarray}
b=  \rho \frac{\pa}{\pa \rho} (1/u \Db P + 1/\ut D \Pt) + h.c.
\end{eqnarray}
where
\begin{eqnarray}
\{b, \Om\} \simeq -i \frac{\pa}{\pa X}, && \{b,\De\} =0.
\end{eqnarray}
The idea is that the old b-field expression acts on the
$G_\De$-ghost number one states, and its hermitian conjugate acts
on the dual ghost number 0 state. There is a similar modification
for $h$.

\section{A Partition function}

Another way to find the spectrum is through a zero-momentum
partition function, following the approach of Berkovits and
Nekrasov\cite{Berkovits:2005hy} for the superparticle.  The
Lefschetz trace formula shows how the graded trace over physical
states of a BRST invariant operator is equal to that over all
states, as explained in chapter 14 of \cite{Henneaux:1992ig}. By a
suitable choice of operator, we can get information about the
spectrum without calculating it directly. This reads as
\begin{equation}
\chi(t)=\text{Tr}_{\text{Phys}} (-)^F t^K = \text{Tr}_{\text{All}}
(-)^F t^K
\end{equation}
where $t$ is a free parameter, $F$ is the fermion number and K is
a BRST invariant number operator
\begin{eqnarray}
F= \theta^a \frac{\pa}{\pa \theta^a} -
 \frac{\pa}{\pa \rho} \rho, && K =  \theta^a \frac{\pa}{\pa \theta^a} + u^a \frac{\pa}{\pa u^a} +
2  \frac{\pa}{\pa \rho} \rho.
\end{eqnarray}
The trace over all states splits up into a product of separate
traces over $\theta$, $u$ and $\rho$ respectively
\begin{equation}
\chi(t) =  (1 -t)^2 (1-t)^{-2}(-t^2(1-t^{-2})) = 1-t^2
\end{equation}
The $1$ corresponds to a bosonic scalar with K-number 0, and the
$-t^2$ refers to a fermionic scalar with K-number 2. These are the
$C$ and $\Ct$ respectively in equation \eqref{eq:wavefnc}.


\section{Equivalence to a simple bosonic model}

Consider the following system with nilpotent BRST charge
\begin{eqnarray}
\Om = \eta \frac{\pa}{\pa X}, 
\end{eqnarray}
where $\eta$ is a fermionic ghost, and $X$ is as before. This
happens to be the simplest non-trivial BRST system. A general
wavefunction
\begin{equation}
\psi = C(X) + \eta \Ct(X)
\end{equation}
in the cohomology $H(\Om)$ obeys
\begin{eqnarray}
\pa_X C(X) = 0, && \de\Ct = \pa_X G(X),
\end{eqnarray}
for some function $G(X)$. Since this matches the state cohomology
of our model as seen in section \ref{sec:state_cohomology}, the
two systems are equivalent. To summarize, our supersymmetric
system with bosonic ghosts is exactly equivalent to a bosonic one
with a fermionic ghost.



\section{The Topological String}

If we take a direct sum of three copies of the above bosonic model
labelled by $i=1,2,3$, and complexify the $X^i$'s, we end up with
the particle version of the Topological string B-model. The BRST
charge is $\Om= \eta^i \pa_{X^i}$ which corresponds to the
exterior derivative. Doing the same for our equivalent toy model,
the BRST charge is given by $\Om = u^{a i} D_{a i}$, where each
variable now has an $i$ index. We can move from world-line to
world-sheet to get the full string theory, in the manner of the
pure spinor string.



As an exercise, we construct the topological B-model open
superstring with a flat background metric. The action
\begin{equation}
S=\int{d^2 z (\half \pab X^i \pa X^{\jb} \eta_{i \jb} + \half \pab
X^{\ib} \pa X^{j} \eta_{\ib j}+ \pab \theta^{a i} \pi_{z a i} +
\pab u^{a i} v_{z a i} +  \pab \xi_i  \rho_z^i + \text{right
movers}) }
\end{equation}
is in the style of the pure spinor string, where $\pi_z$, $v_z$
and $\xi$ are left-moving conjugate momenta to $\theta$, $u$ and
$\rho$ respectively. The BRST charges are now
\begin{eqnarray}
\Om = \oint{dz u^{ai} d_{z a i}}, && \De = \sum_i \oint{dz
\rho^i_z u^{ai}u_{ai}},
\end{eqnarray}
where $d_{z ai}=\pi_{z ai} - i(\pa X_{i} -\frac{i}{2} \theta_{bi}
\pa \theta^{bi}) \theta_{ai}$ is the stringified covariant
derivative, with no summation implied over $i$.

The respective central charge contributions from $X^i$,
$\theta^{ai}$, $u^{ai}$ and $\rho^i_z$ are 6, - 12, +12 and -6. It
is gratifying that they sum to zero as expected.


\begin{acknowledgments}
I would like to thank Louise Dolan and Pierre Vanhove for useful
discussions.
\end{acknowledgments}

\bibliography{referencelibrary}
\bibliographystyle{hunsrt}
\end{document}